\documentclass[journal, 10pt]{IEEEtran}

\usepackage{url}

\usepackage{graphicx}
\usepackage{amsmath, amssymb, amsfonts}
\usepackage{mathtools}
\usepackage{bm}
\usepackage[detect-mode=false, mode=text]{siunitx}
\usepackage{xcolor}

\usepackage[linesnumbered,ruled,noend,vlined]{algorithm2e}
\SetArgSty{textrm}
\SetNoFillComment
\DontPrintSemicolon

\usepackage{cleveref}
\usepackage{lipsum}
\usepackage{booktabs}

\usepackage[sort&compress,square,numbers]{natbib}

\usepackage{jabbrv}
\usepackage{comment}

\DefineSpuriousJournalWord{The}

\DefineJournalException{IEEE/ACM Transactions on Audio Speech and Language Processing}{IEEE/ACM Trans. ASLP}
\DefineJournalException{IEEE Transactions on Audio Speech and Language Processing}{IEEE Trans. ASLP}

\DefineJournalException{Journal of the Acoustical Society of America}{J. ASA}
\DefineJournalException{The Journal of the Acoustical Society of America}{J. ASA}

\DefineJournalException{Proceedings of the IEEE International Conference on Acoustics, Speech and Signal Processing}{ICASSP}
\DefineJournalException{Proceedings of the 2022 IEEE International Conference on Acoustics, Speech and Signal Processing}{ICASSP}
\DefineJournalException{Proceedings of the 2017 IEEE International Conference on Acoustics, Speech and Signal Processing}{ICASSP}

\DefineJournalException{Proceedings of the Annual Conference of the International Speech Communication Association}{Interspeech}

\DefineJournalException{Proceedings of the 45th International Congress and Exposition on Noise Control Engineering}{Internoise}
\DefineJournalException{Proceedings of the 46th International Congress and Exposition on Noise Control Engineering}{Internoise}
\DefineJournalException{Proceedings of the 49th International Congress and Expo on Noise Control Engineering}{Internoise}
\DefineJournalException{Proceedings of the 51st International Congress and Expo on Noise Control Engineering}{Internoise}

\DefineJournalException{Proceedings of the International Congress on Acoustics}{ICA}

\DefineJournalException{Proceedings of the Conference on Empirical Methods in Natural Language Processing}{EMNLP}

\DefineJournalException{Proceedings of the IEEE/CVF Conference on Computer Vision and Pattern Recognition}{CVPR}

\DefineJournalException{Proceedings of the 3rd International Conference on Learning Representations}{ICLR}
\DefineJournalException{Proceedings of the 6th International Conference on Learning Representations}{ICLR}

\DefineJournalException{Proceedings of the Conference on Neural Information Processing Systems}{NeurIPS}

\DefineJournalException{Proceedings of the 38th International Conference on Machine Learning}{ICML}

\DefineJournalException{Proceedings of the 34th AAAI Conference on Artificial Intelligence}{AAAI}

\Crefname{algorithm}{Algo.\@}{Algo.\@}
\Crefname{figure}{Fig.\@}{Fig.\@}

\begin{document}\sloppy

\title{%
Autonomous In-Situ Soundscape Augmentation\\via Joint Selection of Masker and Gain
}

\author{%
    Karn~N.~Watcharasupat, \IEEEmembership{Student~Member, IEEE}, 
    Kenneth~Ooi, \IEEEmembership{Student~Member, IEEE}, 
    Bhan~Lam, \IEEEmembership{Member, IEEE}, 
    Trevor~Wong, \IEEEmembership{Student~Member, IEEE}, 
    Zhen-Ting~Ong,
    and
    Woon-Seng~Gan,~\IEEEmembership{Senior~Member,~IEEE}
\thanks{This work was supported by the National Research Foundation and Ministry of National Development, Singapore, under the Cities of Tomorrow~R\&D Program (COT-V4-2020-1); the Google Cloud Research Credits program (GCP205559654); and the AWS Singapore Cloud Innovation Center. 
Any opinions, findings and conclusions or recommendations expressed in this material are those of the authors and do not reflect the view of the National Research Foundation and the Ministry of National Development, Singapore.
}
\thanks{All authors are with the 
School of Electrical \& Electronic Engineering, Nanyang Technological University (NTU), Singapore. Email: \{karn001, wooi002, bhanlam, trevor.wong, ztong, \mbox{ewsgan}\}@ntu.edu.sg. 
Research protocols
used in this work have been approved by the NTU Institutional Review Board (IRB-2020-08-035).}
}

\markboth{Submitted to Signal Processing Letters}
{Watcharasupat \MakeLowercase{\textit{et al.}}: Autonomous In-Situ Soundscape Augmentation via Joint Selection of Masker and Gain}
\maketitle

\begin{abstract}
    The selection of maskers and playback gain levels in an in-situ soundscape augmentation system is crucial to its effectiveness in improving the overall acoustic comfort of a given environment. 
    Traditionally, the selection of appropriate maskers and gain levels has been informed by expert opinion, which may not be representative of the target population, or by listening tests, which can be time- and labor-intensive. Furthermore, the resulting static choices of masker and gain are often inflexible to dynamic real-world soundscapes. 
    In this work, we utilized a deep learning model to perform joint selection of the optimal masker and its gain level for a given soundscape. The proposed model was designed with highly modular building blocks, allowing for an optimized inference process that can quickly search through a large number of masker-gain combinations. In addition, we introduced the use of feature-domain soundscape augmentation conditioned on the digital gain level, eliminating the computationally expensive waveform-domain mixing process during inference, as well as the tedious gain adjustment process required for new maskers. The proposed system was evaluated on a large-scale dataset of subjective responses to augmented soundscapes with 442 participants, with the best model achieving a mean squared error of $\num[detect-weight]{0.122}\mathbf{\pm}\num[detect-weight]{0.005}$ on pleasantness score, validating the ability of the model to predict combined effect of the masker and its gain level on the perceptual pleasantness level. The proposed system thus allows in-situ or mixed-reality soundscape augmentation to be performed autonomously with near real-time latency while continuously accounting for changes in acoustic environments.
\end{abstract}


\IEEEpeerreviewmaketitle

\section{Introduction}

\IEEEPARstart{M}{itigation} of urban noise pollution is a multifaceted problem where elimination or reduction of noise sources is often impractical \cite{Kang2019TowardsIndices}. Unlike traditional noise control approaches of reducing the sound pressure level (SPL) or acoustic energy, the soundscape approach presents a more holistic and human-centric strategy based on perceptual acoustic comfort \cite{vanRenterghem2019InteractiveSetting}. One such technique commonly termed \textit{soundscape augmentation} involves \textit{adding} ``wanted'' sound(s) into the acoustic environment so as to ``mask'' noise and improve the overall perceptual acoustic quality, with promising results in virtual and real settings across typical urban noise source types~\cite{%
Hong2020EffectsQuality,             % natural 
Coensel2011EffectsNoise,            % natural -> traffic
% Leung2016OnNoise,                   % natural -> traffic
Hao2016AssessmentEnvironment,       % bird -> traffic
% Zhao2020EffectPark,                 % bird -> restore
% Lawton2020NatureExperience,         % nature 
% Lacey2020ThreeSpaces,               % ??
% VanRenterghem2020InteractivePark,   % natural
Trudeau2020ASpace%                  % water
% Cai2019EffectNoise,                 % water
% Lugten2018ImprovingVegetation%      % water
}. 

However, the selection of ``maskers'' and playback levels, both in research and in practice, has traditionally been arbitrary \cite{Leung2016OnNoise}, expert-guided \cite{Hong2021ASounds}, or based on post-hoc analysis~\cite{vanRenterghem2019InteractiveSetting}. Such selection processes, unless part of the study design, are not only disproportionately time- and labor-intensive, but also inflexible to dynamic real-world soundscapes. As a result, it is often unlikely that these static choices of maskers and playback levels are able to consistently achieve optimal acoustic comfort. A recent work \cite{DePessemier2022EnhancingSoundscape} utilized a human-in-the-loop design for public soundscape augmentation, but the process required active participation and resulting soundscapes became somewhat personalized. Existing autonomous models targeting augmentation also tend to focus on only a few maskers of the same `class', e.g. birds \cite{Jahani2021AnAreas}. A number of other prediction models of soundscape experience have also been developed \cite[see][]{Lionello2020ASoundscapes}. However, most of these were linear or shallow non-linear models fitted on limited number of data points and/or only taking psychoacoustic indicators, rather than the spectral representations, as inputs. As a result, semantic or contextual meanings of sound sources are usually lost in these models. To address these limitations, \cite{Ooi2022ProbablyAugmentation} presented one of the first attempts at adaptive augmentation, via a masker selection system with a probabilistic deep learning model that predicts distributions of ``pleasantness'' for augmented soundscapes 

Although the probabilistic perceptual attribute predictor (PPAP) model in \cite{Ooi2022ProbablyAugmentation} presented a proof of concept for automatic neural soundscape augmentation, the model was not made for in-situ deployment by design. The model in \cite{Ooi2022ProbablyAugmentation} requires a premixed track of augmented soundscape as an input. The mixing process was done at pre-calculated soundscape-to-masker ratios (SMR) relative to known in-situ SPLs of the unaugmented soundscape tracks. Thus the model is blind to masker gain level, a crucial playback parameter for in-situ deployment. Moreover, requiring a mixture of the base soundscape recording with a masker as an input adds unnecessary complexity to the overall system, as all candidate maskers would require a lookup table of the digital gain required for playback at a specific SMR, and each candidate masker-SMR pair would require computing an augmented soundscape track.

In this work, we reformulated the PPAP model to allow for a more compute- and bandwidth-efficient system in real-world in-situ or mixed-reality deployment. The proposed model decouples the base soundscape, the masker, and the gain level from one another at the input stage, and introduces separate feature extractor branches for the base soundscape and the masker. The ``augmentation'' process only occurs in the feature space instead of the waveform space by using the digital gain level of the masker, which is independent from the in-situ SPL, as a conditioning input. Attention mechanism is then used to ``query'' the distribution of the target attribute from the base soundscape feature, the masker feature, and the gain-conditioned feature. The proposed method results in complexity reduction in multiple components of the system. First, the raw waveform of the soundscape need not be sent to the inference engine to be mixed with the maskers, as the less bandwidth-consuming spectral data can be used. 
% For cloud inference, this means significant reduction in data egress rate from the edge. 
Second, the masker features can be precomputed independently from the gain level, reducing the inference time and overall latency. 
Third, attribute prediction on multiple gain levels of the same masker can be more quickly performed without recomputing the augmented soundscape, or even the soundscape and masker features. Lastly, addition of new maskers to the system can be done without any need for the creation of a lookup table.

% The rest of the paper is organized as follows. \Cref{sec:proposed} discusses the proposed method and \Cref{sec:data} discusses the dataset. \Cref{sec:expt} discuss the experimental setup and the results. Finally, \Cref{sec:concl} concludes the paper.

\section{Proposed Method}\label{sec:proposed}

% \subsection{Preliminaries}
Consider a digital soundscape recording $\bm{s}_i[n] \in [-1, 1]^C$, with $C\ge1$ channels and sample index $n$. Denote the in-situ A-weighted equivalent SPL, $L_\text{A,\,eq}$, in dBA, of $\bm{s}_i$ by $a_i$. For brevity, all mentions of SPL will refer to the 30-second $L_\text{A,\,eq}$. Suppose all soundscapes are recorded by the same in-situ recording setup, and assuming approximate linearity, the SPL-to-digital level ratio (SPDR), which is the the ratio of the relative SPL in the linear scale to the digital full scale (DFS) of the setup, can be considered roughly constant and denoted by $d_0$.
% , whose unit is $\si{\pascal}/\left(p_0 \cdot \text{DFS}\right)$, where $p_0=\SI{20}{\micro\pascal}$ is the reference SPL. 
Since all soundscape inputs originate from the same recording device, the in-situ SPL information is implicitly embedded in the input spectrogram data.

Consider a single-channel masker $m_j[n]\in[-1, 1]$, for which we do not assume a common recording setup. 
In practice, this is due to the maskers being typically sourced from different recording systems from the in-situ recording setup, and often via open-content providers. As such, the SPDR of each masker is often unknown and cannot be assumed to be constant. Let the SPDR of masker $m_j$ be $d_j$. Under the approximate linearity assumption\footnote{This was verified by exhaustive gain measurement of all maskers via \cite{Ooi2021AutomationHead}.}, a masker $m_j$ can be normalized to the same SPDR as the soundscape recording by a scaling of $d_0/d_j$.
If both $d_j$ and the in-situ SPL $a_i$ are both known, soundscape augmentation with a particular SMR, in dBA, can proceed without gain adjustment. However, $d_j$ is often unknown in practice, and the gain adjustment required to obtain accurate values of $g_j(\cdot)$ typically requires specialized equipment, such as a recording setup with an artificial head, and can be cost-prohibitive. 
Therefore, $g_j$ is used explicitly as a conditioning input, eliminating the need for the SPL-dependent SMR or masker gain adjustment.

\subsection{Model}

\begin{figure}[t]
    \centering
    \includegraphics[width=0.6\columnwidth]{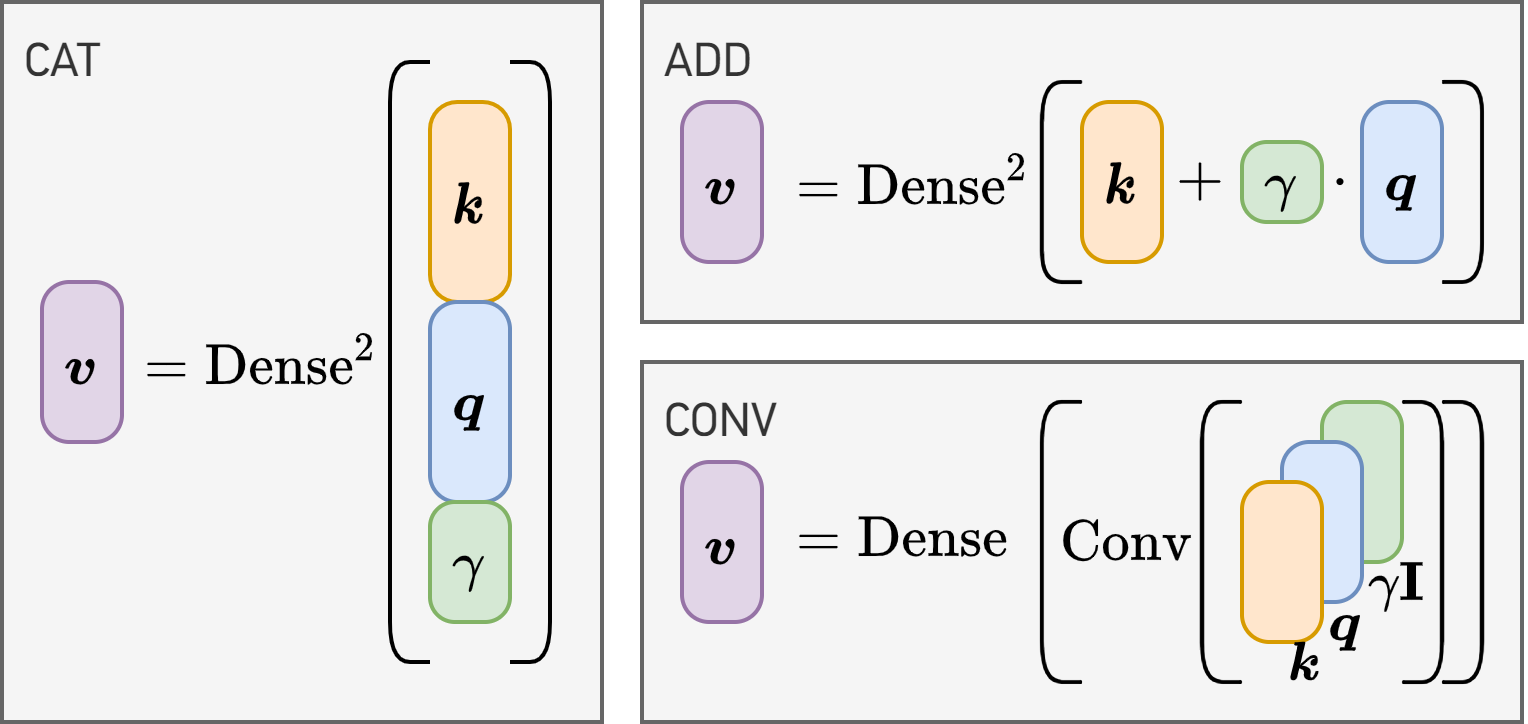}
    \caption{Illustration of the three feature augmentation methods}
    \label{fig:aug}
\end{figure}

\begin{figure}
    \centering
    \includegraphics[width=\columnwidth]{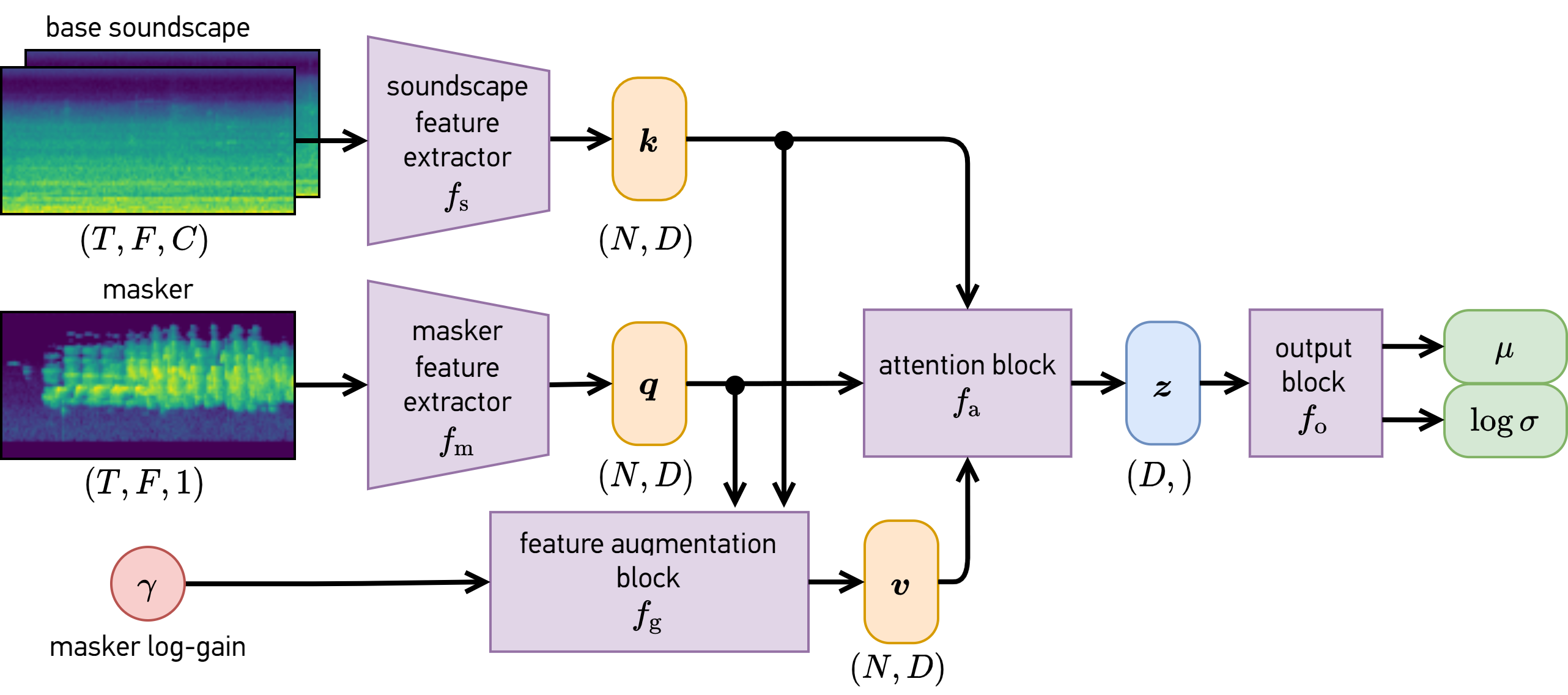}
    \caption{Overview of the PPAP model architecture}
    \label{fig:model}
\end{figure}

Let $\mathbf{S}_i\in\mathbb{R}^{T\times F \times C}$ and $\mathbf{M}_i\in\mathbb{R}^{T\times F \times 1}$ be the log-mel spectrogram of soundscape $\bm{s}_i$ and masker $m_j$, respectively, with $T$ spectrogram time frames and $F$ mel bins. Here, $C=\num{2}$, $F=\num{64}$, and $T=644$, corresponding to \SI{30}{\second} of signal sampled at \SI{44.1}{\kilo\hertz} with a short-time Fourier transform window size of \num{4096} samples and \SI{50}{\%} overlap. The soundscape and masker embeddings are respectively calculated by the feature extractors $f_\text{s}$ and $f_\text{m}$, such that
\begin{equation}
    \bm{k}_{i} = f_\text{s}\left(\mathbf{S}_i\right) \in \mathbb{R}^{N\times D},\
    \bm{q}_{j} = f_\text{m}\left(\mathbf{M}_j\right) \in
    \mathbb{R}^{N\times D},
\end{equation}
where $D=\num{128}$ is the embedding dimension, and $N=\num{20}$ is the number of feature time frames. The feature extractors follow a convolutional design, a well-tested method for information retrieval from spectrogram data \cite{Hershey2017CNNClassification, Kong2020PANNs:Recognition}. Each feature extractor consists of 5 convolutional blocks, each containing a $\num{3}\times\num{3}$ convolutional layer; batch normalization; dropout; Swish activation; and a $\num{2}\times\num{2}$ average pooling. The blocks contain 16, 32, 48, 64, and 64 output channels in this order. 

Under the query-key-value (QKV) model of attention \cite{Vaswani2017AttentionNeed}, the perceptual attribute prediction process can be viewed as using a masker to query a mapping from the unaugmented soundscape (keys) to the augmented soundscape (values). Instead of performing soundscape augmentation in the audio or spectral domain, we consider performing the ``augmentation'' in the embedding domain. This is done via a simple mapping, conditioned on the masker digital gain, such that
\begin{equation}
    \bm{v}_{i, j, g}
    = f_\text{g}\left(\bm{k}_{i},\bm{q}_{j},\gamma\right) \in \mathbb{R}^{N\times D},
\end{equation}
where $g$ is the querying gain level, $\gamma=\log_{10}g$,
and $f_\text{g}$ is a gain-conditioned augmentation layer. We experimented with three different implementations for $f_\text{g}$, namely,
\begin{align}
    f_\text{g}^\textsc{(cat)}(\bm{k}_{i},\bm{q}_{j},\gamma)
    &= \operatorname{Dense}^2\left(\bm{k}_{i}\shortmid\bm{q}_{j}\shortmid\gamma\mathbf{I}^{N\times 1}\right),\\
    f_\text{g}^\textsc{(add)}(\bm{k}_{i},\bm{q}_{j},\gamma)
    &= \operatorname{Dense}^2\left(\bm{k}_{i}+\gamma\bm{q}_{j}\right),\\
    f_\text{g}^\textsc{(conv)}(\bm{k}_{i},\bm{q}_{j},\gamma)
    &= \operatorname{Dense}\left(\operatorname{Conv}\left(\bm{k}_{i}\shortparallel\bm{q}_{j}\shortparallel\gamma\mathbf{I}^{N\times D}\right)\right),
\end{align}
where $\shortmid$ is tensor concatenation, $\shortparallel$ is tensor stacking, $\operatorname{Dense}\colon\mathbb{R}^{N \times \bullet}\mapsto\mathbb{R}^{N \times D}$ is a dense layer with $D$ output units, $\operatorname{Conv}\colon\mathbb{R}^{N \times D \times 3}\mapsto\mathbb{R}^{N \times D}$ is a convolutional layer with $D$ filters with the one-dimensional kernels compressing the stacked dimension into a singleton axis. \Cref{fig:aug} illustrates the feature augmentation process. During training, if the masker $m_j$ is a silent track, we randomize $\gamma\sim\mathcal{N}\left(\upsilon, \zeta^2\right)$, where $\upsilon$ and $\zeta$ are the mean and standard deviation of the log-gains of the training samples with non-silence maskers, to teach the model to ignore gain values for silent maskers. 

With the query, key, and value embeddings, the output embedding can be computed using any QKV attention $f_a$, using
\begin{equation}
    \bm{z}_{i, j, g} = f_\text{a}\left(
        \bm{q}_{j},
        \bm{k}_{i},
        \bm{v}_{i, j, g}
    \right)\in\mathbb{R}^{D}.
\end{equation}
The predicted distribution $\hat{Y}_{i,j,g}$ is then computed by the output block $f_o$ which predicts its parameters $\mu_{i,j,g}$ and $\log \sigma_{i, j, g}$ from $\bm{z}_{i, j, g}$. In this work, we consider three types of attention, namely, additive attention (\textsc{aa}) \cite{Bahdanau2015NeuralTranslate}, dot-product attention (\textsc{dpa}) \cite{Luong2015EffectiveTranslation}, and multi-head attention (\textsc{mha}) \cite{Vaswani2017AttentionNeed} with 4 heads. The use of attention in the fusion block is due to its tested ability in summarizing multiple sequential embeddings that need not originate from the same source or even modality~\cite{Jaegle2021Perceiver:Attention, Xuan2020Cross-modalLocalization}. \Cref{fig:model} illustrates the proposed architecture.

\subsection{Loss Function}

As with \cite{Ooi2022ProbablyAugmentation}, we consider human subjective responses to be inherently random, both in the sense that a participant is a sample of the population, and that the response given by each participant during the listening test is also a random sample of their inherently non-deterministic perception. As such, the label $y_{i,j,g}$ for the target attribute is considered an observation of a random variable $Y_{i,j,g}$ representing the \textit{distribution} of the target attribute, whose true distribution is unknown. Using the maximum likelihood formulation, the optimization objective of the PPAP model can be given by
\begin{equation}
    \textstyle\max_{\mathbf{\Theta}} \sum_{i,j,g} \mathcal{L}_{\hat{Y}_{i,j,g}}\left(y_{i,j,g}\right) \label{eq:optim}
\end{equation}
where $\Theta$ is the parameters of the model, and $\mathcal{L}_{X}(x)$ is the log-density of $X$ evaluated at $x$. Modeling the subjective response to each acoustic scene by an independent normal distribution, the optimization \eqref{eq:optim} translates to the loss function
\begin{equation}
    \textstyle\mathcal{J} = \frac{1}{|\mathcal{T}|}\sum_{i,j,g} \frac{1}{2}\left[({y_{i,j,g}-\mu_{i,j,g}})/{\sigma_{i,j,g}}\right]^2 + \log \sigma_{i,j,g},
\end{equation}  
where $\mathcal{T}$ is the training batch.

\subsection{Optimized Inference}

Although the model sees batches of different soundscape-masker-gain samples during training, the model sees only one base soundscape at time during in-situ inference, while going through multiple maskers and gain levels to find the most suitable masker-gain pair for the particular soundscape. Therefore, significant runtime reduction can be made by eliminating duplicate computation. Denote the time required to perform $f_\text{s}$, $f_\text{m}$, and $f_\text{g}\circ f_\text{a} \circ f_\text{o}$ by $\tau_\text{s}$, $\tau_\text{m}$, and $\tau_\text{g+a+o}$, respectively. Let $\eta_\text{m}$ be the number of maskers to query per soundscape and $\eta_\text{g}$ be the number of gain values to query per masker. A naive batching scheme would require a total runtime of $\eta_\text{g}\eta_\text{m}(\tau_\text{s}+\tau_\text{m}+\tau_\text{g+a+o})$. Instead, by optimizing the query process, as shown in \Cref{alg:query}, the total runtime can be reduced to $\tau_\text{s}+\eta_\text{m}\tau_\text{m}+\eta_\text{g}\eta_\text{m}\tau_\text{g+a+o}$. Additionally, for a fixed masker bank, precomputing the masker features $\bm{q}_{j}$ will further reduce the runtime to to $\tau_\text{s}+\eta_\text{g}\eta_\text{m}\tau_\text{g+a+o}$. 

\begin{algorithm}[t]
\small
\caption{Optimized Inference}\label{alg:query}

\KwIn{Soundscape recording $\bm{s}_i$}
% \tcp{on edge node}
$\mathbf{S}_i = \operatorname{LogMelSpec}(\bm{s}_i)$ \;
% \;
% \tcp{on inference engine from here}
$\bm{k}_{i} = f_\text{s}\left(\mathbf{S}_i\right)$\;
\ForEach{masker spectrogram $\mathbf{M}_j$}{
    $\bm{q}_{j} = f_\text{m}\left(\mathbf{M}_j\right)$ \tcp*{can be precomputed}
    % $\bm{q}_{j} = f_\text{m}\left(\operatorname{LogMelSpec}(m_j)\right)$
    \ForEach{gain $g$}{
        $\bm{v}_{i, j, g}
    = f_\text{g}\left(\bm{k}_{i}, \bm{q}_{j}, \gamma\right)$\;
        $Y_{i,j,g} = f_\text{o}\circ f_{a}\left(
        \bm{q}_{j},
        \bm{k}_{i},
        \bm{v}_{i, j, g}
    \right)$\;
    }
}
\KwOut{Predicted pleasantness distributions $\{Y_{i,j,g}\}$}
\end{algorithm}

\section{Dataset} \label{sec:data}

This work utilized an early version of the ARAUS dataset \cite{Ooi2022ARAUS:Soundscapes} with \num{18564} data points from the first 442 participants (42 stimuli each). Each data point represents the subjective responses of a participant to an augmented soundscape. 
% \subsection{Stimuli}
The augmented soundscapes were made in 5 folds by adding \num{30}-second ``maskers'' to \num{30}-second binaural recordings of soundscapes from the Urban Soundscapes of the World (USotW) database \cite{DeCoensel2017UrbanMind}. All recordings used were sampled at \SI{44.1}{\kilo\hertz}. Further details can be found in \cite{Ooi2022ARAUS:Soundscapes, Ooi2022ProbablyAugmentation}.
Addition of the maskers to the soundscapes were made at specific SMRs of \{\num{-6}, \num{-3}, 0, 3, 6\} dBA.
% , where SMR is defined by the ratio of the SPL of the soundscape to that of the masker. 
Since the masker data were sourced from online repositories Freesound \cite{Font2013FreesoundDemo} and Xeno-canto \cite{Planque2008Xeno-canto:Song}, they required gain adjustment. A lookup table of the digital gain $g_j(l)$ required to amplify the $j$th masker to a specific SPL of $l$ dBA was created by exhaustively gain-adjusting the track on a dummy head 
% (GRAS 45BB KEMAR Head \& Torso), 
to SPL values between \SI{46}{dBA} and \SI{83}{dBA} inclusive, in \SI{1}{dBA} steps. The gain adjustment process was semi-automated via \cite{Ooi2021AutomationHead} using the same headphones and soundcards as those of the participants. Since the in-situ SPL of the base soundscapes are usually non-integral, the digital gain is interpolated from the lookup table using
\begin{equation}
    g_j(l) = \text{cg}_j[\text{round}(l)]\cdot10^{(l - \text{round}(l))/20},
\end{equation}
where $\text{cg}_j[\lambda]$ is the calibrated gain for masker $j$ at $\lambda$ dBA~SPL.

% \subsection{Subjective Responses}\label{ssec:subj}

All participants listened to the calibrated augmented soundscapes on a pair of circumaural headphones 
% (Beyerdynamic Custom One Pro), 
powered by an external sound card. 
% (Creative SoundBlaster E5). 
After listening to each augmented soundscape, participants responded to the \num{8}-item questionnaire based on the ISO/TS 12913-2:2018 standard \cite{InternationalOrganizationforStandardization2018ISO/TSRequirements}:
\begin{quote}
    \textit{To what extent do you agree or disagree that the present surrounding sound environment is [...]?}
\end{quote}
where \textit{[...]} is one of \textit{pleasant}, \textit{eventful}, \textit{uneventful}, \textit{chaotic},  \textit{vibrant}, \textit{calm}, \textit{annoying}, and \textit{monotonous}. 
Each questionnaire item has five-point Likert options from ``Strongly disagree''~(1) to ``Strongly agree'' (5). The ground-truth label of ISO Pleasantness \cite{InternationalOrganizationforStandardization2019ISO/TSAnalysis} for each augmented soundscape is then calculated as per \cite[eq. (4)]{Ooi2022ProbablyAugmentation}. Note that \textit{eventful} and \textit{uneventful} ratings were not used for ISO Pleasantness calculation. 

\section{Experiments} \label{sec:expt}

\begin{figure}[t]
    \centering
    \includegraphics[width=0.8\columnwidth]{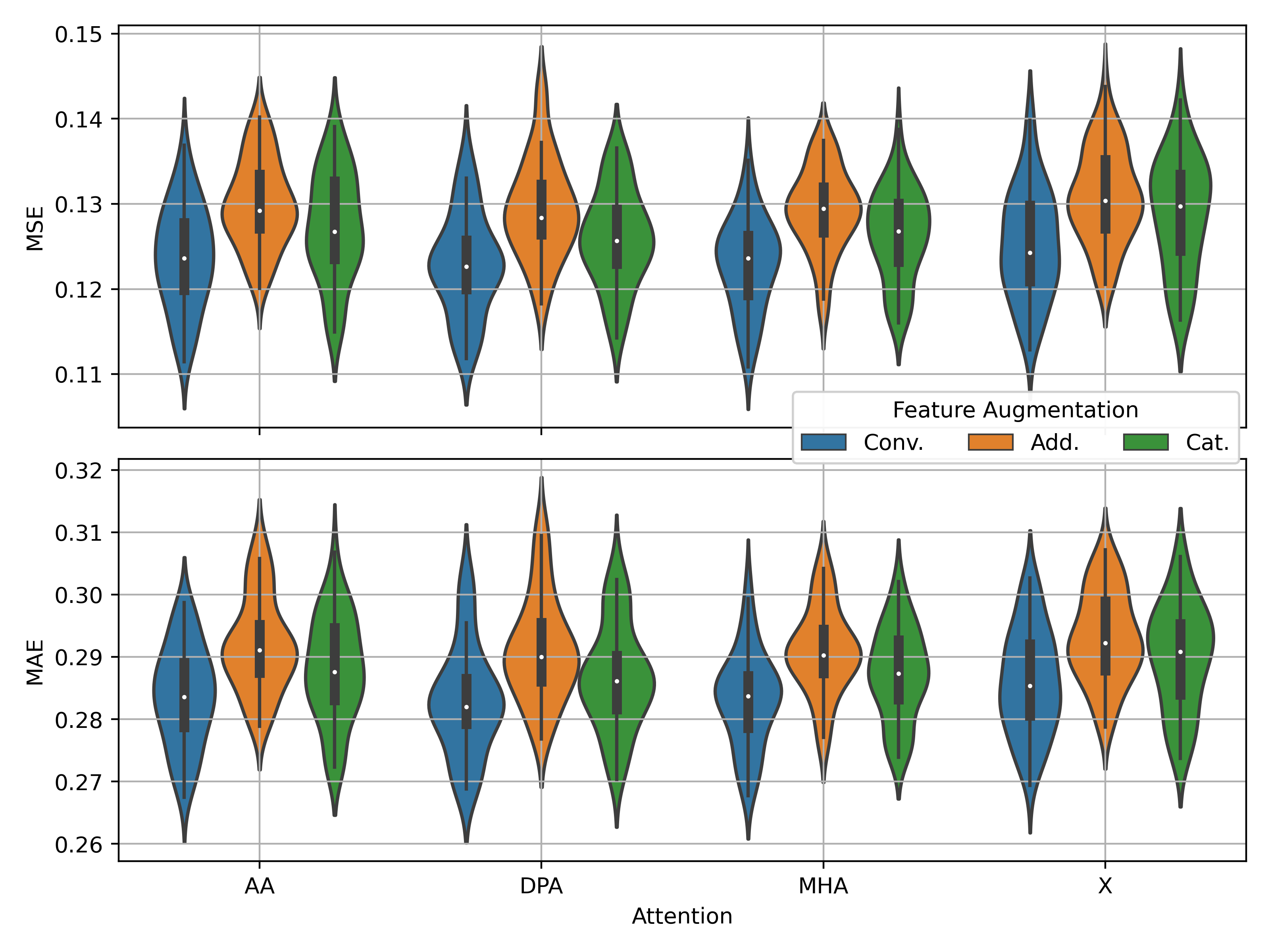}
    \vspace{-\baselineskip}
    \caption{Violin plots of the validation MSE and MAE of the pleasantness prediction for each attention type.}
    \label{fig:results}
\end{figure}

As an ablation model, we also experimented with a ``pass-through'' attention block, denoted by \textsc{x} in \protect\Cref{fig:results} and defined~by
\begin{equation}
    \operatorname{PassThroughAttention}\left(
        \bm{q}_{j},
        \bm{k}_{i},
        \bm{v}_{i, j, g}
    \right) = \bm{v}_{i, j, g}.
\end{equation} 
For all model types in this paper, each model was trained in a 5-fold cross-validation manner for the same 10 seeds per validation fold, totalling 50 runs per model type. Each model was trained for up to 100 epochs using an Adam optimizer with a learning rate of \num{5e-5}. 
% The learning rate is halved if the validation mean square error (MSE) does not decrease for at least 5 epochs, and the training is early-stopped if the validation MSE does not decrease for at least 10 epochs. 
 \Cref{fig:results} shows the distribution of the prediction errors in terms of MSE and mean absolute error (MAE) for each attention and augmentation type.

Across all attention types, the \textsc{conv} augmentation performed the best in terms of both MSE and MAE, followed by \textsc{cat} and \textsc{add}. We believe that this could be attributed to the \textsc{conv} method encouraging feature alignment between $\bm{k}_i$ and $\bm{q}_j$, and the kernel filtering allowing a more flexible augmentation operation in the feature space compared to the other two. It can also be seen that using a QKV attention generally allows better performance compared to the no-attention model. However, it appears that the type of attention mechanism does not significantly affect model performance, as long as some form of attention is used to couple the queries, keys, and values. Despite \textsc{aa} and \textsc{dpa} only having a single attention head, the four-head \textsc{mha} does not seem to result in a noticeable improvement in prediction error. It should also be noted that the choice of feature augmentation block seems to more significantly impact the accuracy of the model, as seen by the \textsc{conv} models without attention generally performing better than the \textsc{add} models with attention.

Despite the model never seeing the augmented soundscape audio presented to the participants, the model with best average MSE (\textsc{dpa}, \textsc{conv}) at $\num{0.122}\pm\num{0.005}$ is on par with the result in  \cite{Ooi2022ProbablyAugmentation}, which used premixed augmented soundscapes as inputs. This demonstrates the effectiveness of the feature augmentation technique in emulating soundscape augmentation in the feature space instead of the waveform space. 

\begin{figure}[t]
    \centering
    \includegraphics[width=0.8\columnwidth]{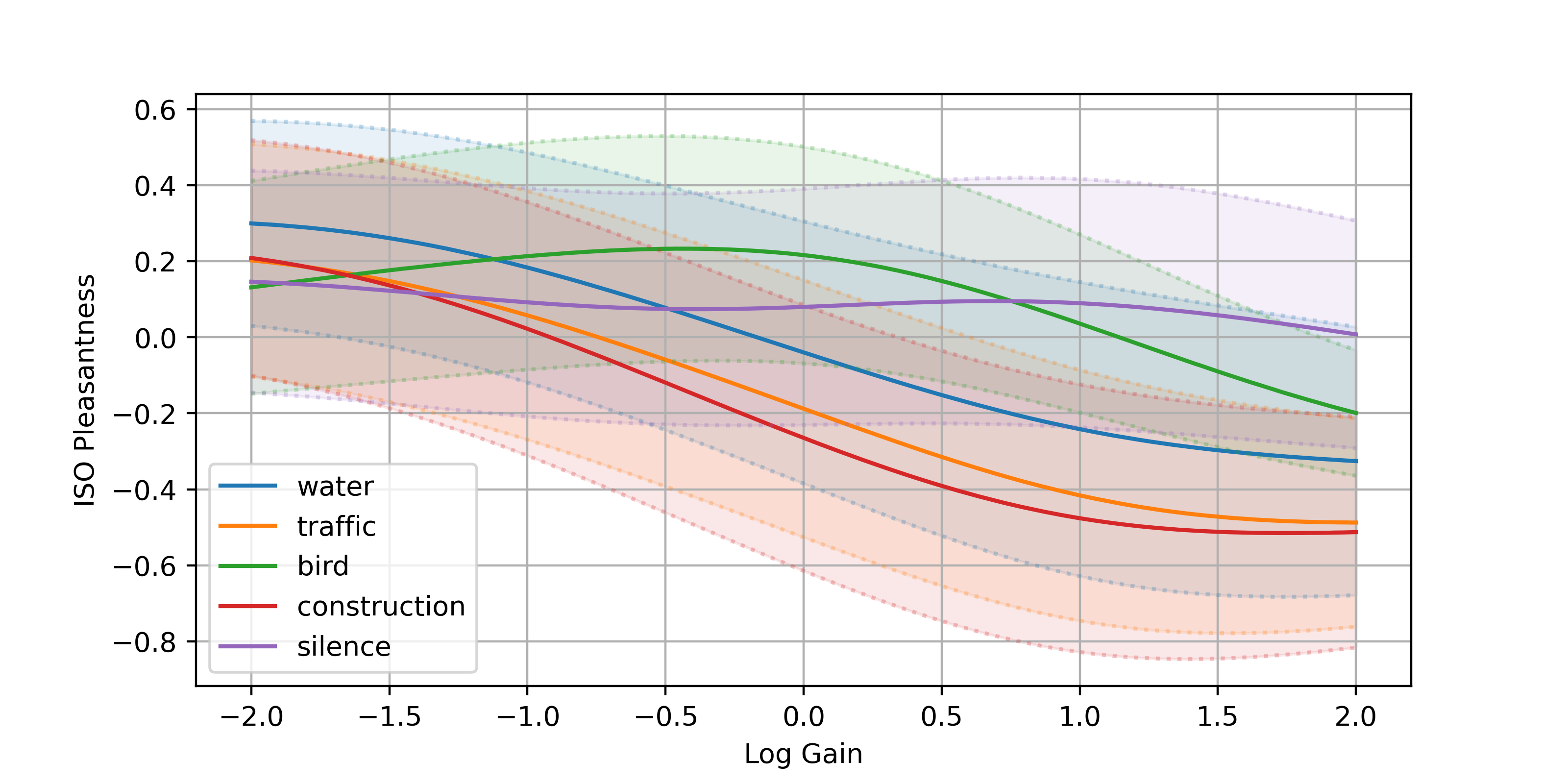}
    \vspace{-1em}
    \caption{Gain interpolation using an \textsc{aa} model with \textsc{conv} augmentation (seed 5, validation fold 2). All maskers here were pre-adjusted to \SI{65.0}{dBA} at $\gamma=0$. 
    % The base soundscape has an in-situ SPL level of \SI{65.32}{dBA}. The base soundscape and all maskers are unseen by the model during training. The solid line represents the mean of the predicted distribution. The shaded region represents the pleasantness level up to one predicted standard deviation above and below the predicted mean.
    }
    \label{fig:gain}
\end{figure}

In \Cref{fig:gain}, we demonstrate the gain-aware nature of the model by querying the ISO Pleasantness distribution across \num{256} values of $\gamma\in[\num{-2}, \num{2}]$ for different types of maskers. A silent `masker' track was also included to test the model's ability to consider the joint effects of the masker and its gain level. Although not perfectly so, it can be seen that the model can discern when the masker is silent (i.e., the soundscape is unaugmented), outputting a roughly constant pleasantness prediction regardless of $\gamma$. For non-silent maskers, the model seems to be aware of maskers' nature.  In line with \cite{Hao2016AssessmentEnvironment, Zhao2020EffectPark, Lam2019EvaluationDevice, Ong2019PredictionReality, Hong2021ASounds}, the model predicted increasing pleasantness as the gain of the bird masker increases until it is close to the ambience level ($\gamma\approx\num{-0.5}$), followed by a drop in pleasantness below that of the unaugmented soundscape. A similar effect is also seen with the water masker, with the pleasantness level being above that of the unaugmented soundscape up to $\gamma\approx\num{-0.5}$, although the pleasantness level starts dropping at lower gain than the bird masker due to the more continuous nature of the water sound \cite{Jeon2010PerceptualSounds, Hong2021ASounds}. For known unpleasant maskers, such as traffic and construction, the model correctly outputs increasingly unpleasant predictions as the gain level increases. 

The proposed model is currently being deployed in-situ with a speaker-array setup \cite[see][]{Wong2022DeploymentAugmentation}. Although the current version of the model only considers one masker track at a time and does not allow for a composition of multiple maskers, the modular design could be adapted for multiple maskers. In future work, the modular design could also allow additional control or conditioning parameters, such as environmental context and audience demographics, to be included in the masker selection process, similar to the post-hoc model in \cite{Mitchell2021InvestigatingApproach}.

\section{Conclusion} \label{sec:concl}

In this work, we proposed an improved probabilistic perceptual attribute predictor model that allows gain-aware prediction of subjective responses to augmented soundscapes. The proposed model decouples the masker and soundscape feature extraction and emulates soundscape augmentation in the feature space, eliminating the need for computationally expensive mixing in the waveform domain. Additionally, the model was reformulated to consider the digital gain level of the masker, instead of the soundscape-to-masker ratio, allowing the use of new maskers from various sources without the need for time-consuming gain adjustment. The modular design of the model allows for significant feature reuse and pre-computation during inference time, reducing the overall latency and computational resources required in deployment. Using a large-scale dataset of \SI{18}{K} subjective responses from 442 participants, we demonstrated the ability of the model to accurately predict ISO Pleasantness in relation to soundscapes, maskers, and gains. 

\bibliographystyle{IEEEbiba}
\bibliography{references}

\end{document}